\documentclass[fleqn,usenatbib]{mnras}
\usepackage{newtxtext,newtxmath}
\usepackage[T1]{fontenc}

\DeclareRobustCommand{\VAN}[3]{#2}
\let\VANthebibliography\thebibliography
\def\thebibliography{\DeclareRobustCommand{\VAN}[3]{##3}\VANthebibliography}

\def\revisionchris#1{{\textcolor{black}{{#1}}}}

\newcommand{\HI}{H{\sc~i}}
\newcommand{\HII}{H{\sc~ii}}

\newcommand{\HeII}{He{\sc~ii}}

\newcommand{\angstrom}{\mbox{\normalfont\AA}}

\usepackage{graphicx}	% Including figure files
\usepackage{amsmath}	% Advanced maths commands
\title[Neutral islands]{The effect of reionization on direct measurements of the mean free path}

\author[J. Roth et al.]{Joshua T. Roth$^{1}$\thanks{jroth021@ucr.edu}, Anson D'Aloisio$^{1}$, Christopher Cain$^{2,1}$, Bayu Wilson$^{1}$, Yongda Zhu$^{1}$, \newauthor and George D. Becker$^{1}$
\\ 
$^{1}$Department of Physics and Astronomy, University of California, Riverside, CA 92521, USA
\\
$^{2}$School of Earth and Space exploration, Arizona State University, Tempe, AZ 85281, USA}

\begin{document}
\label{firstpage}
\pagerange{\pageref{firstpage}--\pageref{lastpage}}
\maketitle

\begin{abstract}
Recent measurements of the ionizing photon mean free path (MFP) based on composite quasar spectra may point to reionization ending at $z<6$.  These measurements are challenging because they rely on assumptions about the proximity zones of the quasars.  For example, some quasars might have been close to neutral patches where reionization was still ongoing (``neutral islands"), and it is unclear how they would affect the measurements. We address this question with mock MFP measurements from radiative transfer simulations.  We find that, even in the presence of neutral islands, our mock MFP measurements agree to within $30~\%$ with the true spatially averaged MFP in our simulations, which includes opacity from both the ionized medium and the islands.  The inferred MFP is sensitive at the $<~50\%$ level to assumptions about quasar environments and lifetimes for realistic models. We demonstrate that future analyses with improved data may require explicitly modeling the effects of neutral islands on the composite spectra, and we outline a method for doing this.    Lastly, we quantify the effects of neutral islands on Lyman-series transmission, which has been modeled with optically thin simulations in previous MFP analyses. Neutral islands can suppress transmission at $\lambda_{\rm rest}<912$~\AA\ significantly, up to a factor of 2 for $z_{\rm qso}=6$ in a plausible reionization scenario, owing to absorption by many closely spaced lines as quasar light redshifts into resonance.  However, the suppression is almost entirely degenerate with the spectrum normalization and thus does not significantly bias the inferred MFP. 
\end{abstract}

\bigskip

\begin{keywords} dark ages, reionization, first stars --- intergalactic medium --- QSOs: absorption lines
\end{keywords}

%%%%%%%%%%%%%%%%%%%%%%%%%%%%%%%%%%%%%%%%%%%%%%%%%%

%%%%%%%%%%%%%%%%% BODY OF PAPER %%%%%%%%%%%%%%%%%%

\section{Introduction}
\label{sec:intro}

Cosmic reionization marks the emergence of the first \HI-ionizing sources and the transition of the intergalactic medium (IGM) from neutral to highly ionized.  The bulk of this transition likely took place between $z = 12$ and $6$, with evidence from the Ly$\alpha$ forest of $z>5$ quasars (QSOs hereafter) suggesting that the tail of reionization extended to $z \approx 5.3$~\citep{Kulkarni2019,Keating2019,Nasir2020,Qin2021,2022MNRAS.514...55B,Zhu2022}. Another line of evidence for such a late reionization process comes from recent measurements of the mean free path (MFP) of ionizing photons at $5 < z < 6$ \citep{Becker2021,Bosman2021b,Gaikwad2023,2023arXiv230804614Z}.  In this paper, we point out and assess physical effects which could, in principle, be important for the interpretation of some of these measurements.    

\citet[][B21]{Becker2021} measured the MFP by stacking a sample of rest-frame QSO spectra and fitting the aggregate shape of the stack at wavelengths $\lambda < 912$ \AA.  They model the composite spectrum flux as
\begin{equation}
f^{\rm obs}_{\lambda}= f^{\rm SED}_{\lambda} \exp(-\tau^{\rm Lyman}_{\rm eff}) \exp(-\tau^{\rm LyC}_{\rm eff}) + f_0, 
\label{eq:total_fobs}
\end{equation}
where $f_{\lambda}^{\rm SED}$ is the intrinsic spectral energy distribution (SED) of the QSO, $f_0$ is a zero-point parameter, and $\tau^{\rm Lyman}_{\rm eff}$ and $\tau^{\rm LyC}_{\rm eff}$ are the effective optical depths from Lyman-series and Lyman-continuum (LyC) absorption, respectively.\footnote{The $f_0$ parameter accounts for uncertainties in the quasar spectra zero-point estimates and subtractions.} The B21 method is essentially the same as the one introduced by \citet{Prochaska2009}, but with the key improvement that the QSO proximity effect is included in the model.  Doing so is critical at redshifts approaching $z=6$ because during reionization the MFP is expected to become smaller than the proximity radius of the QSOs.  In this case, neglecting the proximity effect can bias the MFP measurement high \citep{DAloisio2018}. 

The short MFP measured by B21 at $z = 6$, $\lambda_{912}^{\rm mfp} = 3.57_{-2.14}^{+3.09}$ $h^{-1}$cMpc, indicates a factor of $10$ evolution in the IGM opacity in the $200$ Myr between $z=5.1$ and $z=6$ when combined with pre-existing $z < 5.2$ measurements \citep{Worseck2014}. {Subsequently, \citet[][Z23]{2023arXiv230804614Z} updated the analysis with an expanded set of QSOs from the XQR-30 program \citep{DOdorico2023}, and new Keck/ESI observations, to provide the first measurements based on QSO composites at $z = 5.6$ and $5.3$.}  Their results are consistent with the rapid evolution found by B21.    

The rapid evolution has been interpreted as a signature of the final stages of reionization \citep{Cain2021,Garaldi2022,Lewis2022}. However, if reionization were ongoing at $5 < z < 6$, some of the QSO spectra in the B21 and Z23 analyses might intersect last remaining neutral regions in the IGM, termed neutral islands.  These neutral islands could affect the MFP measurements in several ways.  For one, LyC absorption by the islands could affect the shape of the composite spectra and thus spoil the accuracy of the B21/Z23 functional form for $\exp(-\tau^{\rm LyC}_{\rm eff})$, which was derived neglecting potential effects from neutral islands.  Secondly, even if their functional form remains a good approximation, neutral islands could affect the relationship between the excess ionizing background produced by the QSO and the IGM opacity. In B21/Z23, this relationship is parameterized by an ansatz for the Lyman limit absorption coefficient, 
\begin{equation}
\kappa_{912} = \kappa^{\rm bg}_{912}\left( \frac{\Gamma}{\Gamma_{\rm bg}} \right)^{-\xi},
\label{eq:kappa_scaling}
\end{equation}
where $\Gamma$ is the local \HI\ photoionziation rate, and $\kappa^{\rm bg}_{912}$ and $\Gamma_{\rm bg}$ are the background absorption coefficient and photoionzation rate, respectively, i.e. the spatially averaged values in the absence of the QSO proximity effect. To choose plausible values for the power-law index $\xi$, B21/Z23 drew on guidance from theoretical calculations for optically thick absorbers in the {\it ionized} IGM.  But neutral islands could alter the response of $\kappa_{912}$ to $\Gamma$ in a way that results in different effective values of $\xi$.  Lastly, high-order Lyman-series absorption by neutral islands could alter the shape of the $\exp(-\tau^{\rm Lyman}_{\rm eff})$ term.  If not modeled, these effects could introduce systematic errors into the MFP analysis. The goal of this work is to quantify them and discuss their implications for the B21/Z23 method.  

Recently, \citet{Gaikwad2023} reported independent measurements of the MFP obtained by comparing simulations to the measured distribution of Ly$\alpha$ opacity fluctuations (see also Davies et al. in prep.).  Although compatible with the B21 and Z23 measurements (given large uncertainties), the best-fit values from these forest-based measurements at $z=6$ are a factor of $\approx 2$ higher than the central values reported by the former.  Such differences might be explained if neutral islands act to bias the QSO-composite-based measurements low, further motivating the present work. {During the preparation of this manuscript, we became aware of another analysis by Satyavolu et al. in prep., which uses a suite of radiative transfer simulations to forward model the $\exp(-\tau^{\rm LyC}_{\rm eff})$ term (rather than the parametric model of B21) to measure the MFP. They also find a short value of $\lambda_{912}^{\rm mfp} = 4.28_{-1.57}^{+2.95}$ $h^{-1}$cMpc, consistent with the B21/Z23 measurements.  }  

\begin{figure*}
	\centering
	\includegraphics[scale=0.23]{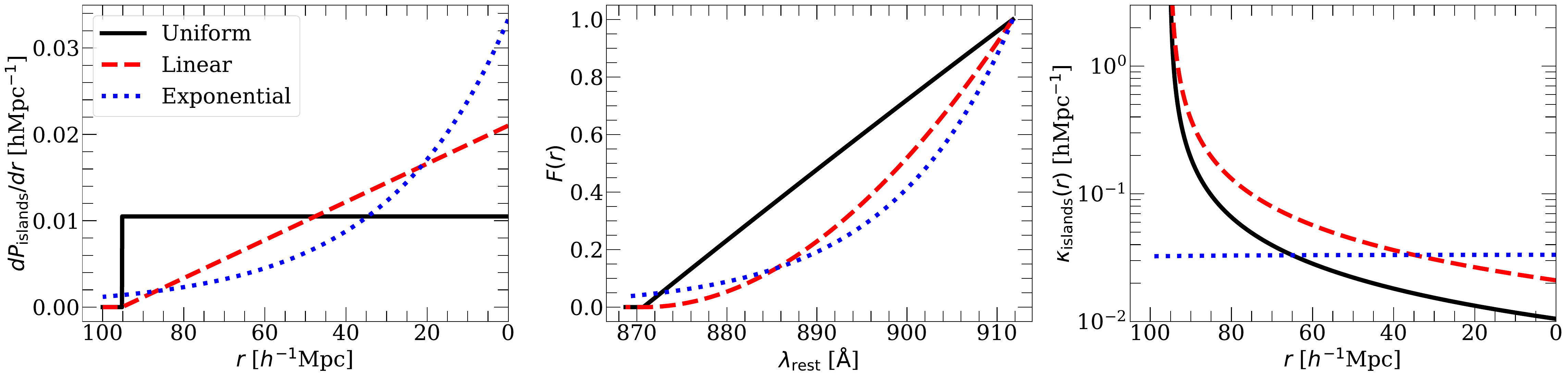}
	\caption{Model transmission spectra and absorption coefficients spectra intersecting only neutral islands.  Left: $dP_{\rm islands}/dr$ for uniform (black solid), linear (red dashed), and exponential (blue dotted) distributions.  Middle: transmission spectra for each model as given by Eq.~\ref{eq:dFdr}.  Right: absorption coefficient $\kappa(r)$ (Eq.~\ref{eq:kappa_island_r}) for each model.  In the exponential model, $\kappa(r)$ is constant and the islands behave effectively like a spatially uniform contribution to the IGM opacity.  }
	\label{fig:kappa_island_example}
\end{figure*}
 
This paper is organized as follows.  In \S \ref{sec:B21method} we give a brief review of the B21/Z23 method.  In \S\ref{sec:analytical}, we present an analytic model which we use to gain intuition.  In \S\ref{sec:methods}, we describe our procedure for generating mock QSO spectra using radiative transfer simulations.  We present our main results in \S\ref{sec:results} and \S \ref{sec:LyN}. The former isolates the effects of neutral islands in the $\exp(-\tau^{\rm LyC}_{\rm eff})$ term, while that latter incorporates the $\exp(-\tau^{\rm Lyman}_{\rm eff})$ term as well.  We conclude in \S\ref{sec:conc}.  Throughout this work, we assume the following cosmological parameters: $\Omega_m = 0.305$, $\Omega_{\Lambda} = 1 - \Omega_m$, $\Omega_b = 0.048$, $h = 0.68$, $n_s = 0.9667$ and $\sigma_8 = 0.82$, consistent with~\cite{Planck2018} results. All distances are in co-moving units unless otherwise specified.

\section{Summary of the B21/Z23 method}
\label{sec:B21method}

We begin with a brief summary of the B21/Z23 method for measuring the MFP. As described in \S \ref{sec:intro}, a composite spectrum is constructed by averaging the rest-frame spectra of a sample of QSOs in a given redshift bin.  Blue-ward of $\lambda = 912$ \AA, the composite is modeled with Eq. (\ref{eq:total_fobs}).  The effective LyC opacity  is assumed to take the form,
\begin{equation}
\label{eq:taueff_LyC}
\tau^{\rm LyC}_{\rm eff}(z_{912},z_{\rm qso}) = \frac{c~(1+z_{\rm 912})^{2.75}}{H_0 \Omega_m^{1/2}} \int^{z_{\rm qso}}_{z_{912}} dz' \kappa_{\rm 912}(z')(1+z')^{-5.25},
\end{equation}
for a photon emitted at $z_{\rm qso}$ that redshifts to 912 \AA\ at $z=z_{912}$ \citep{Prochaska2009}.  Here $\kappa_{912}(z)$ denotes the absorption coefficient at 912 \AA\ at redshift $z$. Eq. (\ref{eq:taueff_LyC}) assumes that the absorption coefficent is proportional to the \HI\ photoionization cross section, $\kappa(\nu) \propto \sigma_{\rm HI}(\nu)$, where $\nu$ is frequency and the cross section is approximated as $\sigma_{\rm HI} = \sigma_0 (\nu/\nu_{912})^{-2.75}$. {(See e.g. discussion in \citealt{2023MNRAS.tmp.3051T})}.  

Previous analyses at $z<5$ took $\kappa_{\rm 912}$ to be a constant when fitting Eqs. (\ref{eq:total_fobs}) and (\ref{eq:taueff_LyC}) to a composite spectrum \citep{Prochaska2009, Worseck2014}.  An important improvement introduced by B21, which allows the measurements to be extended to higher redshift, is the inclusion of the QSO proximity effect in $\kappa_{\rm 912}(z)$.  Motivated by analytic models \citep{MiraldaEscude2000,Furlanetto2005} and simulations of the IGM opacity \citep{McQuinn2011, Park2016, DAloisio2020, Nasir2021}, the B21/Z23 model assumes that the absorption coefficient has the power-law dependence on $\Gamma$ given by Eq. (\ref{eq:kappa_scaling}). The total $\Gamma$ is the sum of contributions from the background and the QSO, $\Gamma(r) = \Gamma_{\rm bg} + \Gamma_{\rm qso}(r)$, where the latter introduces a dependence on distance $r$ from the QSO.  Rewriting Eq. (\ref{eq:kappa_scaling}) in terms of the two contributions yields  
\begin{equation}
\kappa_{912}(r) = \kappa^{\rm bg}_{912}\left[1 + \frac{\Gamma_{\rm qso}(r)}{\Gamma_{\rm bg}} \right]^{-\xi}.
\label{eq:kappa_scaling2}
\end{equation}
Note that Eq. (\ref{eq:kappa_scaling2}) is an implicit equation that must be solved numerically because $\Gamma_{\rm qso}(r)$ is attenuated by the optical depth from $\kappa_{912}$ at distance less than $r$. 

B21/Z23 express the QSO ionizing luminosity, in terms of $\Gamma_{\rm bg}$, using $R_{\rm eq}$, the formal distance at which $\Gamma_{\rm qso} = \Gamma_{\rm bg}$ not including absorption by intervening matter or cosmological redshifting.  QSO luminosities are typically reported at rest-frame 1450 \AA, $L_{1450}$. For a broken power-law QSO continuum, 
\begin{equation}
  L_{\rm \nu}=\begin{cases}
    \nu^{-\alpha_\nu^{\rm UV}} & 912~\angstrom < \lambda < 1450 ~\angstrom \\
    \nu^{-\alpha_\nu^{\rm ion}} & \lambda < 912~\angstrom,
  \end{cases}
\end{equation}
the QSO luminosity at 912 \AA\ is $L_{\rm 912} = L_{\rm 1450} (\nu_{\rm 912}/\nu_{\rm 1450})^{-\alpha^{\rm UV}_\nu}$, and it can be easily shown that 
\begin{equation}
R_{\rm eq} = \left[ \frac{L_{912}~\sigma_0}{4 \pi \Gamma_{\rm bg} (\alpha_\nu^{\rm ion}+2.75)} \right]^{1/2}, 
\end{equation}
where, again, the \HI\ photoionization cross section is approximated as $\sigma_{\rm HI} = \sigma_0 (\nu/\nu_{912})^{-2.75}$. Note also that the QSO SED blue-ward of 912 \AA, i.e. the prefactor in Eq. (\ref{eq:total_fobs}), is $f^{\rm SED}_{\lambda} = f_{912}( \lambda / 912$ \AA $)^{-\alpha^{\rm ion}_\nu}$, so $f_{912}$ parameterizes the overall normalization at 912 \AA. The B21/Z23 model for the LyC term thus consists of five parameters, $f_{\rm 912}$, $f_0$, $\kappa^{912}_{\rm bg}$, $R_{\rm eq}$, and $\xi$. For a given $L_{\rm 1450}$, $R_{\rm eq}$ is estimated using Ly$\alpha$ forest measurements of $\Gamma_{\rm bg}$ and measurements of $\alpha_\nu^{\rm UV}$ and $\alpha_\nu^{\rm ion}$ from composite QSO spectra (see B21 and Z23 for more details). Unfortunately, the current data do not allow for an informative simultaneous fit with $\xi$.  Instead, B21 and Z23 explore a range of fixed values $\xi \in [0.33,1]$ motivated by the aforementioned theoretical studies. This leaves 3 free parameters to be fit: $f_{\rm 912}$, $f_0$, $\kappa^{912}_{\rm bg}$. 

To model the $\exp(-\tau^{\rm Lyman}_{\rm eff})$ term in the fitting function (\ref{eq:total_fobs}), B21/Z23 compute $\tau^{\rm Lyman}_{\rm eff}$ from skewers traced through the 40-1024 run of the Sherwood simulation suite \citep{2017MNRAS.464..897B}.  The simulated Ly$\alpha$ optical depths are re-scaled to match the observed mean Ly$\alpha$ forest flux compiled over $z=3.5-6$ from \citet{Becker2013} and \citet{Bosman2018}.  The total effective optical depth, $\tau^{\rm Lyman}_{\rm eff}$, is obtained by summing the effective opacities from the first 39 Lyman-series lines. Note that once the Ly$\alpha$ optical depths are re-scaled, the effective optical depths of the remaining 38 lines are completely determined by the simulation.  In \S \ref{sec:LyN}, we will return to the topic of the $\exp(-\tau^{\rm Lyman}_{\rm eff})$ term. 

In the B21/Z23 anaylses the measured MFP is defined to be the proper distance to $\tau_{\rm eff}^{\rm LyC}(z_{912},z_{\rm qso}) = 1$.\footnote{Operationally, using the $\kappa^{912}_{\rm bg}$ from the fit, they find the wavelength $\lambda$ at which $\tau_{\rm eff}^{\rm LyC} = 1$ in the absence of the QSO proximity zone.  They then compute the  $z= \lambda(1+z_{\rm qso})/ 911.76 \angstrom -1$, and calculate the proper distance between $z_{\rm qso}$ and $z$.} At $z>5$, the redshifts of interest here, the MFP measured in this way is to a very good approximation $1/\kappa^{912}_{\rm bg}$. A key point is that $\kappa^{912}_{\rm bg}$ parameterizes the average opacity of the IGM absent the QSO proximity effect; hence $\lambda^{912}_{\rm mfp}$ corresponds formally to the characteristic attenuation length of the background IGM, not of the immediate neighborhoods of the QSOs.  In what follows, one of our central goals is to test whether the B21/Z23 method probes the MFP of the background IGM even when neutral islands are present.

\section{Analytic model \& insights}
\label{sec:analytical}

\subsection{Neutral island model}

We now present an analytic model for how neutral islands enter the $\exp(-\tau^{\rm LyC}_{\rm eff})$ term of the B21 method.  This model affords some insight into our radiative transfer (RT) results presented below.  \revisionchris{A photon emitted by a QSO at wavelength $\lambda < 912$ will travel a distance $r(\lambda)$ before red-shifting past $912 \text{\AA}$.  In the absence of absorption from the ionized IGM, the probability that this photon will not encounter a neutral island within distance $r$ (the mean transmission $F$) is
\begin{equation}
    \label{eq:dFdr}
    F(\lambda) = 1 - \int_{0}^{r(\lambda)} dr' \left(\frac{dP_{\rm islands}}{dr'}\right)
\end{equation}
where $\frac{dP_{\rm islands}}{dr}$ is the distribution of distances from the QSO to the nearest neutral island.  This distribution is simply the ionized bubble size distribution (BSD).\footnote{Specifically, we use the ``mean free path'' definition of the bubble size distribution introduced in~\cite{Mesinger2007}.}  }
 
We define the neutral island opacity to be
{\begin{equation}
    \label{eq:kappa_island_r}
    \kappa_{\rm islands}(r) \equiv -\frac{1}{F(r(\lambda))}\frac{dF}{dr}
\end{equation}}
Eq.~(\ref{eq:kappa_island_r}) gives the mean opacity from islands as a function of distance from the start of a random sightline.  \revisionchris{Note that this source of effective opacity is independent of wavelength at $\lambda < 912 \text{\AA}$, since the MFP through neutral gas is extremely short ($\sim 1-10$ kpc) at all wavelengths of interest here.}  The total IGM absorption coefficient is the sum of contributions from ionized gas and islands: 
{\begin{equation}
    \label{eq:kappatotal}
    \kappa_{912}^{\rm total}(r) = \kappa_{912}^{\rm ionized}(r) + \kappa_{\rm islands}(r).
\end{equation}}

\begin{figure*}
	\centering
	\includegraphics[scale=0.21]{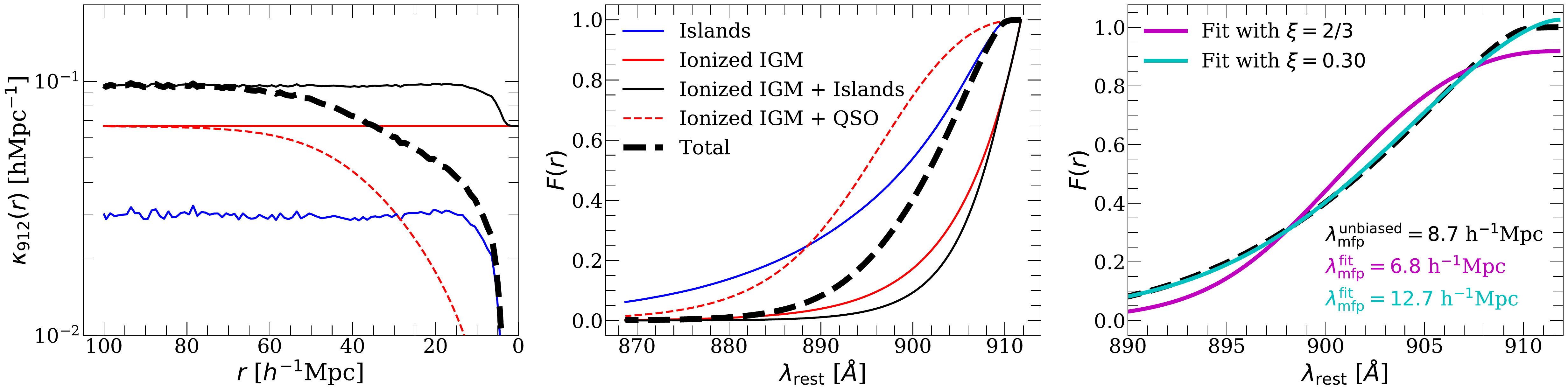}
	\caption{{ Examples of model QSO  spectra containing neutral islands.  The left and middle panels show $\kappa_{912}(r)$ and $F(r)$ for model QSO  spectra. We break down the different contributions to a model QSO  spectrum that contains neutral islands (see legend and main text).  In the right panel, we show that fitting the spectrum containing islands using the B21/Z23 method and assuming the correct $\xi$ for ionized gas (magenta-solid) nonetheless gives a MFP closer to that of the background IGM, i.e. the theoretical value calculated from randomly placed skewers in the IGM, $\lambda^{\rm unbiased}_{\rm mfp}$.  The cyan curve in the right panel shows what happens when we let $\xi$ be a free parameter.  The presence of neutral islands lowers $\xi$ and the fit improves considerably, but the inferred MFP is biased high with respect to $\lambda^{\rm unbiased}_{\rm mfp}$, suggesting that future data sets sensitive enough to constrain $\xi$ may necessitate amending the B21/Z23 model to include possible effects from neutral islands. } }
	\label{fig:qso_example}
\end{figure*}

Fig.~\ref{fig:kappa_island_example} shows examples of $F(r)$ (middle) and $\kappa_{\rm islands}(r)$ (right) for different assumed BSDs (left), where the $x$-axis corresponds to the rest-frame wavelength that redshifts to 912 \AA\ upon traveling a distance $r$.   The solid-black, dashed-red, and dotted-blue curves show uniform, linear, and exponential BSDs, respectively.  For the former two, we assume a maximum bubble radius of $20$ proper Mpc.  In these models $\kappa_{\rm islands}(r)$ increases with $r$, asymptoting vertically at the maximum bubble radius where $F(r)$ goes to $0$.  For the exponential BSD, $\kappa_{\rm islands}(r)$ is constant with distance and acts as a spatially constant contribution to the total IGM opacity.  Below we will see that the exponential model approximates well the $\kappa_{\rm islands}$ extracted from RT simulations of reionization. 

We can add neutral islands to the B21/Z23 model for QSO composite spectra.  Assuming a spatially uniform ionized IGM absorption coefficient $\kappa_{912}^{\rm ion,IGM}$, {the absorption coefficient in ionized gas will follow the form of Eq.~(\ref{eq:kappa_scaling2}). Combining this with Eqs.~(\ref{eq:dFdr}-\ref{eq:kappatotal}) yields}
\begin{equation}
    \label{eq:kappa_total_r}
    \kappa_{912}^{\rm total}(r) = \kappa_{912}^{\rm ion,IGM}\left(1 + \frac{\Gamma_{\rm qso }(r)}{\Gamma_{\rm bg}}\right)^{-\xi} + \frac{dP_{\rm islands}/dr}{1 - \int_{0}^{r} dr' (dP_{\rm islands}/dr')},
\end{equation}
We calculate $\Gamma_{\rm qso }(r)$ by integrating Eq. (8) of B21 and iterating this with Eq.~(\ref{eq:kappa_total_r}) until convergence\footnote{Note that the second term in Eq.~\ref{eq:kappa_total_r} does not include $\Gamma_{\rm qso}$, since neutral islands formally have $\xi = 0$.}.  The initial guess for $\Gamma_{\rm qso }(r)$ is given by Eqs.~(6-7) of B21.  Finally, we note that Eq. (\ref{eq:kappa_total_r}) could be used to extend the B21/Z23 model to include effects of neutral islands, a possibility we plan to explore in future work.  

\subsection{Intuition from model QSO spectra}
\label{sec:intuition}

Fig.~\ref{fig:qso_example} illustrates our model using a BSD at $z=6$ extracted from a cosmological RT simulation of reionization.  The simulation was  run with the adaptive ray-tracing code of \citet{Cain2021, Cain2022b} in a $(200$ $h^{-1}$Mpc$)^3$ volume box and $N = 200^3$ RT grid cells.\footnote{The simulation includes a subgrid model that accounts for the dynamical evolution of un-resolved sinks that set the intergalactic opacity in ionized gas. See \citet{Cain2021, Cain2022b} for more details.}  We populated halos with galactic sources by abundance-matching to the luminosity function of \citet{Finkelstein2019}, and setting \HI-ionizing emissivities proportional to UV luminosity.  {The simulation is calibrated to match the observed mean Ly$\alpha$ forest transmission measured by~\cite{Bosman2021}, as well as the MFP evolution from Z23, as will be described in detail in Gangolli et al. in prep.}    To extract the BSD, we trace sight lines starting on $M_{\rm UV} < -20$ galaxies, which are likely hosts for bright QSOs. The solid-blue curve in Fig.   \ref{fig:qso_example} shows the neutral island opacity as a function of distance from the QSO derived from the extracted BSD. Strikingly, $\kappa_{\rm islands}(r)$ is nearly flat outside of the proximity zone, indicating that the BSD is to a good approximation exponential there, c.f. Eqs. (\ref{eq:dFdr}) and (\ref{eq:kappa_island_r}).  For reference, the volume-weighted mean neutral fraction of the simulation at $z=6$ is $x_{\rm HI} \approx 20 \%$. 

In Fig. \ref{fig:qso_example} we assume a spatially uniform opacity in the ionized IGM, $1/\kappa_{\rm ion,IGM}^{912} = 15$ $h^{-1}$Mpc, which is shown as the solid-red curve in the left panel.  {For this component, we assume $\kappa(\nu) \propto \sigma_{\rm HI}(\nu)$, following the assumption underlying Eq.~(\ref{eq:taueff_LyC}).\footnote{{We could easily extend the model to include the frequency dependence introduced by the column-density distribution in the ionized IGM, but we find that this has little effect on our main results here}.}  }  The solid-black curve shows the net opacity from neutral islands and ionized IGM.  To model the effect of QSO proximity zones, we assume a QSO  UV magnitude of $M_{\rm UV}^{\rm qso } = -27$ and broken power law spectrum with slope $\alpha^{\rm ion}_\nu = 1.5$ and $\alpha^{\rm UV}_\nu = 0.6$, consistent with \citet{2015MNRAS.449.4204L}. Following B21, we adopt $\Gamma_{\rm bg} = 3 \times 10^{-13}$ s$^{-1}$ and $\xi = 2/3$ in Eq. (\ref{eq:kappa_total_r}). The dashed-red curve shows the opacity of the ionized IGM when the QSO proximity zone is included.  Finally, the thick dashed-black curve shows the total opacity with neutral islands and with QSO proximity effect.   

The middle panel shows the resulting LyC spectra. Without the QSO proximity effect, neutral islands have only a modest impact on the spectrum, as can be seen by comparing the solid-red and solid-black curves.\footnote{We have verified this finding numerically using the 1D RT simulation method described in the next section.}  However, with the QSO proximity effect present, the difference induced by the islands is much larger (compare dashed-red and dashed-black curves), since the proximity zones are truncated at the nearest neutral island.   

In the right panel we carry this example further by examining how neutral islands affect the inferred MFP.  {We fit the B21/Z23 model to the LyC spectrum of the total model (dashed-black curve). Specifically, we fit only the $f_{\rm 912} \exp(-\tau^{\rm LyC}_{\rm eff})$ term utilizing Eqs. (\ref{eq:taueff_LyC}) and (\ref{eq:kappa_scaling2}).  We try the fit in two ways: holding $\xi$ fixed and letting it float (see below). In both cases, the free parameters are $f_{912}$ and $\lambda^{\rm fit}_{\rm mfp} = 1/\kappa^{\rm bg}_{\rm 912}$, and $R_{\rm eq}$ is held fixed to its input value.  Based on the discussion in \S \ref{sec:B21method}, we test whether the inferred MFP estimates the actual MFP of the background. For comparison, we calculate the theoretical value of the latter by tracing skewers in random directions and starting at random locations through the neutral island field from the reionization simulation. Inside the ionized bubbles we adopt $1/\kappa_{\rm ion,IGM}^{912} = 15$ $h^{-1}$Mpc.  We compute the theoretical MFP to be $\lambda^{\rm unbiased}_{912} =\int^\infty_0 dr \exp\left[ -\int{\kappa_{912} ~dr}\right] = 8.7~h^{-1}$Mpc, where the term ``unbiased" refers to fact this is the characteristic attenuation scale of the background IGM.  The solid-magenta curve shows the best fit with fixed $\xi = 2/3$ (same as the assumed value for the ionized contribution in Eq. \ref{eq:kappa_total_r}), from which the MFP is inferred to be $\lambda^{\rm fit}_{\rm mfp} = 6.8$ $h^{-1}$Mpc.  This is indeed closer to the background MFP, and considerably shorter than the MFP through the ionized IGM, which, recall, is 15 $h^{-1}$Mpc in this example. At face value, this result comports with the expectation that the B21/Z23 method probes the background MFP. Note, however, that the magenta curve, with fixed $\xi=2/3$, is a poor fit to the model spectrum (black-dashed). The cyan curve shows what happens if we instead let $\xi$ be a free parameter in the fit. In this case the best-fit parameters become $x=0.3$ and $\lambda^{\rm fit}_{\rm mfp} = 12.7 h^{-1}$Mpc. Neutral islands effectively lower the value of $\xi$ by introducing a contribution to the IGM opacity that does not explicitly depend on $\Gamma$ (i.e. $\xi_{\rm islands}=0$), but at the cost of biasing the inferred MFP higher than the background MFP. In what follows, we will explore these issues in greater detail numerically. } 

\section{Numerical Methods}
\label{sec:methods}

We explore these effects now in greater detail using 1D RT post-processed on sight lines extracted from cosmological simulations.  The sight lines are taken from the hydrodynamics simulation originally used in~\cite{Davies2014}, which was run with GADGET-3~\citep{Springel2005} in a box of $L = 3$ $h^{-1}$Mpc and $N = 2\times512^3$ gas and dark matter particles.  The RT code, developed by Cain and Wilson, employs an approach similar to that of \citet{2007MNRAS.374..493B} and \citet{Davies2014}, and will be described in detail in Wilson et. al. in prep.\footnote{We have tested our code against analytic solutions for expansion of an \HII\ region into a uniform medium, and have verified that we get very similar results for the tests shown in Fig. 1 of \citet{Davies2014}} We solve the RT equation along the skewers in 1D assuming a spherically symmetric geometry with a point source at the center.  We track the non-equilibrium ionization states of hydrogen and helium using a backward-difference solution of the chemistry equations.  We use the infinite speed of light approximation, which yields the exact solution for the {\it observed} QSO spectrum along the line of sight \citep{White2003,Shapiro2006}. We employ 40 frequency bins between 1 and 4 Ry. This purposely excludes the ionization of \HeII\ along our sight lines because in this work we would like to isolate and study physical effects related to \HI\ neutral islands and their ionization.  

 We populate our simulated sight lines with neutral islands by tracing random skewers through a separate large-volume 3D simulation of reionization -- the same one described and used in \S \ref{sec:intuition} for the BSD discussed there.  In addition to tracing these skewers from highly biased halos, as described in \S \ref{sec:intuition}, we also consider a set of skewers traced from random locations in the IGM. This is intended to mimic scenarios where at least some QSOs turn on into a locally neutral IGM, and is motivated by our lack of knowledge about high-$z$ QSO hosts.  The neutral islands are grafted onto the hydrodynamic sight lines by setting the neutral fraction to unity at the appropriate comoving distances from the skewer starting points in the 3D simulation.   In ionized regions we apply a spatially uniform photo-ionization rate $\Gamma_{\rm bg}$ and initialize the temperature to $10,000$ K. This approach is a compromise made to achieve higher spatial resolution in the gas properties, compared to the alternative of taking our skewers from a fully self-consistent reionization simulation.

\begin{figure*}
	\centering
	\includegraphics[scale=0.35]{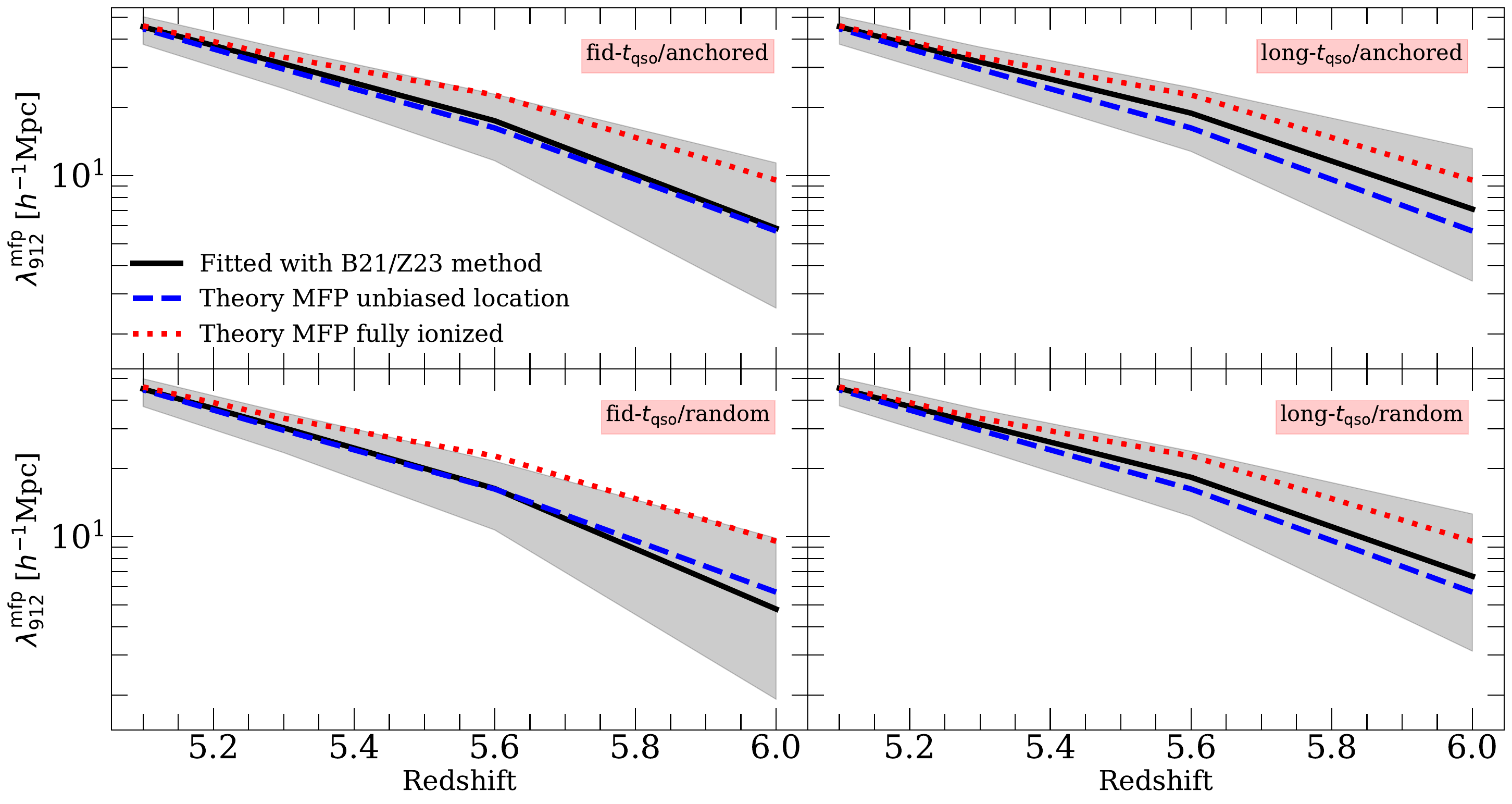}
	\caption{  {Effect of neutral islands on mean free path measurements based on QSO-composites.  We show in each panel (1) the fit, using the method of B21/Z23, to simulations with neutral islands and QSOs assuming $\xi = \xi_{\rm ion}$ (solid black), (2) the theoretical value of the MFP computed along randomly placed sight lines in a reionizing IGM (``unbiased location"; dashed-blue), and (3) the theoretical MFP in a fully ionized IGM (dotted-magenta).  The gray shaded region shows the range of fits for $\xi \in [0.33,1]$.  Each panel shows results for different assumptions about QSO  lifetimes and environments - see main text for details. The fid-$t_{\rm qso}$ and long-$t_{\rm qso}$ refer to assumed distribution of QSO lifetimes, and the anchored and random terminology refers to the initial environment within which the QSO turns on. In the anchored models, the QSOs turn on within already-ionized bubbles during reionization, while in the random model the QSOs may turn on in  fully neutral regions.  }}
	\label{fig:main_results}
\end{figure*}

 The above procedure sets the initial conditions for our 1D RT runs.  The sight lines begin at the center of massive dark matter halos, so we do not include the first 200 proper kpc in our calculations. We turn on a QSO for a lifetime $t_{\rm qso }$, which is drawn from a normal distribution with mean $t_{\rm qso }^{\rm avg}$ and standard deviation $\sigma_{\rm t_{\rm qso }}$. In what follows, we consider two QSO lifetime models: Fiducial $t_{\rm qso}$, with $(t_{\rm qso }^{\rm avg},\sigma_{\rm t_{\rm qso }}) = (5, 1.5)$ Myr, and Long $t_{\rm qso }$ with  $(30, 10)$ Myr. We note that the former gives proximity zone sizes similar to those observed for the XQR-30 sample at $z \sim 6$ \citep{Satyavolu2023}.  The QSOs have $\dot{N}_{\gamma}^{\rm qso } = 1.83 \times 10^{57}$ photons/s, which follows from the adopted $M_{\rm UV}^{\rm qso } = -27$ and the spectral shape (see \S \ref{sec:intuition}). 
 
 Since the $t_{\rm qso}$ are much shorter (up to tens of Myr) than the cosmological time scale over which the gas cools, we employ a computational-time-saving model for the thermal state of the gas, which we find does not alter our main results.  We model the heating from the QSO I-fronts through the neutral islands using the $T_{\rm reion}$ fitting function from \citet{DAloisio2019}.\footnote{In doing so, we correct our I-front speeds for infinite-speed-of-light effects using the procedure described in \citet{DAloisio2019}.}  At relativistic I-front speeds, \citet{2021ApJ...906..124Z} found that non-equilibrium effects suppress $T_{\rm reion}$, so we impose a $T_{\rm reion}$ ceiling of $33,000$ K to account crudely for such effects.  Aside from the I-front heating in the neutral islands, we do not evolve the gas temperatures in time, i.e. the gas outside of neutral islands always remains at $10,000$ K.  We have checked that this approximate model yields temperatures in good agreement with test runs in which the full temperature evolution is solved self-consistently, but reduces the required computational time significantly.      
 
 The results presented in the next section are obtained from a set of $1,000$ sight lines, which are stacked to construct mock LyC composite spectra at $z = 6$, $5.6$, $5.3$, and $5.1$, the redshifts of the Z23 measurements.   In what follows, we will compare mock MFP measurements, obtained with a modified version of the B21 fitting procedure, to the MFP computed directly from the simulations.  For the former, we fit the $f_{\rm 912} \exp(-\tau^{\rm LyC}_{\rm eff})$ term in eq.~(\ref{eq:total_fobs}), letting $f_{912}$ float in the fit, and using Eqs. (\ref{eq:taueff_LyC}) and (\ref{eq:kappa_scaling2}).  (In \S \ref{sec:LyN} we add in the Lyman series term, as described there.) We set $f_0$ to zero because our mock composites do not model zero-point effects. For simplicty, we take the fitted MFP to be $\lambda^{912}_{\rm mfp} = 1/\kappa^{912}_{\rm bg}$, as we find that this definition differs from the B21/Z23 definition (see \S \ref{sec:B21method}) by $<0.2\%$ at $z=6$ and $<3\%$ at $z=5$. We refer to the MFP computed directly from simulations as the ``theory MFP," which we define as $\lambda_{912}^{\rm mfp}= \langle \int^\infty_0 dr \exp[-\tau(r)] \rangle$, where the average is taken over sight lines.\footnote{In evaluating the theory MFP, we take an approach similar to that in \citet{Chardin2015}.}

 The central question we aim to address here is the following. {\it Given that the B21/Z23 fitting method returns the correct (theory) MFP in a fully ionized IGM}, does the presence of neutral islands near the QSOs bias the MFP measurement?  The B21/Z23 fitting procedure requires an estimate for the relationship between $\Gamma(r)$ and the IGM opacity.  This relationship is parameterized by $\xi$ (see Eq.~\ref{eq:kappa_scaling}).  We first run a set of $1,000$ simulations with no QSO or islands to get the theory MFP in the ionized IGM.  Next, we run a set with QSOs (but still without islands) and adjust $\xi$ in the B21/Z23 fitting procedure until the fitted MFP matches the ionized phase theory MFP.  We call this quantity $\xi_{\rm ion}$, and interpret it as the effective value of $\xi$ that properly captures the interaction of the QSO  with the surrounding IGM.\footnote{Note this is not necessarily the same thing as $\xi$ for the photo-ionized IGM without the QSO  present.}  For $z = (6, 5.6, 5.3, 5.1)$, we find $\xi_{\rm ion} = (0.63, 0.58, 0.56, 0.55)$, which we will adopt as our fiducial values. This ensures that the B21/Z23 fitting method returns the theory MFP in a fully ionized IGM, so we can isolate the potential bias introduced by neutral islands.  We will, however, also explore the effect of varying $\xi$. Since we are interested in isolating potential effects of neutral islands, we also assume that $\Gamma_{\rm bg}$ is perfectly known. (We take systemtic errors introduced by an unknown $\Gamma_{\rm bg}$ to be a separate issue outside the scope of this paper.)  We adopt $\Gamma_{\rm bg} = (1.45,3.19,4.04,5.08) \times 10^{-13}$ s$^{-1}$ at $z = (6, 5.6, 5.3, 5.1)$, consistent with values recently measured by \citet{Gaikwad2023} and used in Z23.

\section{The effect of neutral islands on mean free path measurements}
\label{sec:results}

{
In this section we consider the effects of neutral islands through their LyC absorption only.  In the next section we will add in additional effects through the Lyman-series absorption. Figure~\ref{fig:main_results} shows some of our main results.  The panels and curves explore different assumptions about the QSO environments and lifetimes.  In all panels, the solid-black curves show the MFP fitted (using the B21/Z23 model) to our simulated composites. The gray shaded regions show the range of fitted MFPs with $\xi \in [0.33, 1]$, which is the range of $\xi$ assumed in B21/Z23.  
}  

{
 The dashed-blue curves show the theory MFP calculated using the neutral island field from sight lines placed at random locations in the 3D reionization simulation.  Bear in mind in what follows that we do not post-process these sight lines with QSO radiation, and some of these ``unbiased location" sight lines start inside of neutral islands.  The dotted-red curves show the theory MFP calculated in the fully ionized IGM, without neutral islands and without QSO proximity effect. For reference, in the 3D reionization simulation, the volume-weighted mean neutral fractions at $z=6, 5.6, 5.3,$ and $5.1$ are $x_{\rm HI}=22.3\%, 10.0\%, 3.4\%,$ and $0.7\%$, respectively. We see that the fitted and two theory MFPs all converge as the neutral fraction declines, as expected.    
}

The upper left panel of Fig. \ref{fig:main_results} shows results for our fiducial model, with QSO liftetime parameters $(t_{\rm qso }^{\rm avg},\sigma_{\rm t_{\rm qso }}) = (5, 1.5)$ Myr, and initial neutral island distribution taken from sight-lines anchored on $M_{\rm UV} < -20$ galaxies in the 3D reionization simulation. (We call this ``fid-$t_{\rm qso}$~/~anchored.")  Notably, across all redshifts the fit agrees with the theory MFP from unbiased locations within $20\%$. {This is qualitatively consistent with the findings from our simple analytic model in \S \ref{sec:analytical}, and with the expectation that the B21/Z23 method should be probing the background opacity.}  The upper right panel shows the effect of assuming longer QSO life times on average, with $(t_{\rm qso }^{\rm avg},\sigma_{\rm t_{\rm qso }}) = (30, 10)$ Myr. In this case, the QSOs have more time to reionize the neutral islands near them, reducing their effect on the fitted MFP. 

In both of these models, the QSOs turn on inside of pre-existing ionized bubbles created by neighboring galaxies. But it is possible that at least some $z\sim 6$ QSOs turn on inside of neutral regions.  The bottom two panels explore this possibility by using an initial neutral island distribution taken from randomly located sight lines in the 3D reionization simulation.  The bottom-left and bottom-right panel use our fiducial and long $t_{\rm qso}$ models.  The former maximizes the effect of the neutral islands because the QSOs are more likely to start near them, and the shorter $t_{\rm qso}$ increases the likelihood that they will remain in the spectra. A comparison of the four panels highlights the effects of QSO environment and $t_{\rm qso}$. {We find that the inferred MFP depends modestly ($< 50$\% level) on assumptions about the QSO environments and lifetimes.}  

\begin{figure}
    \centering
    \includegraphics[scale=0.290]{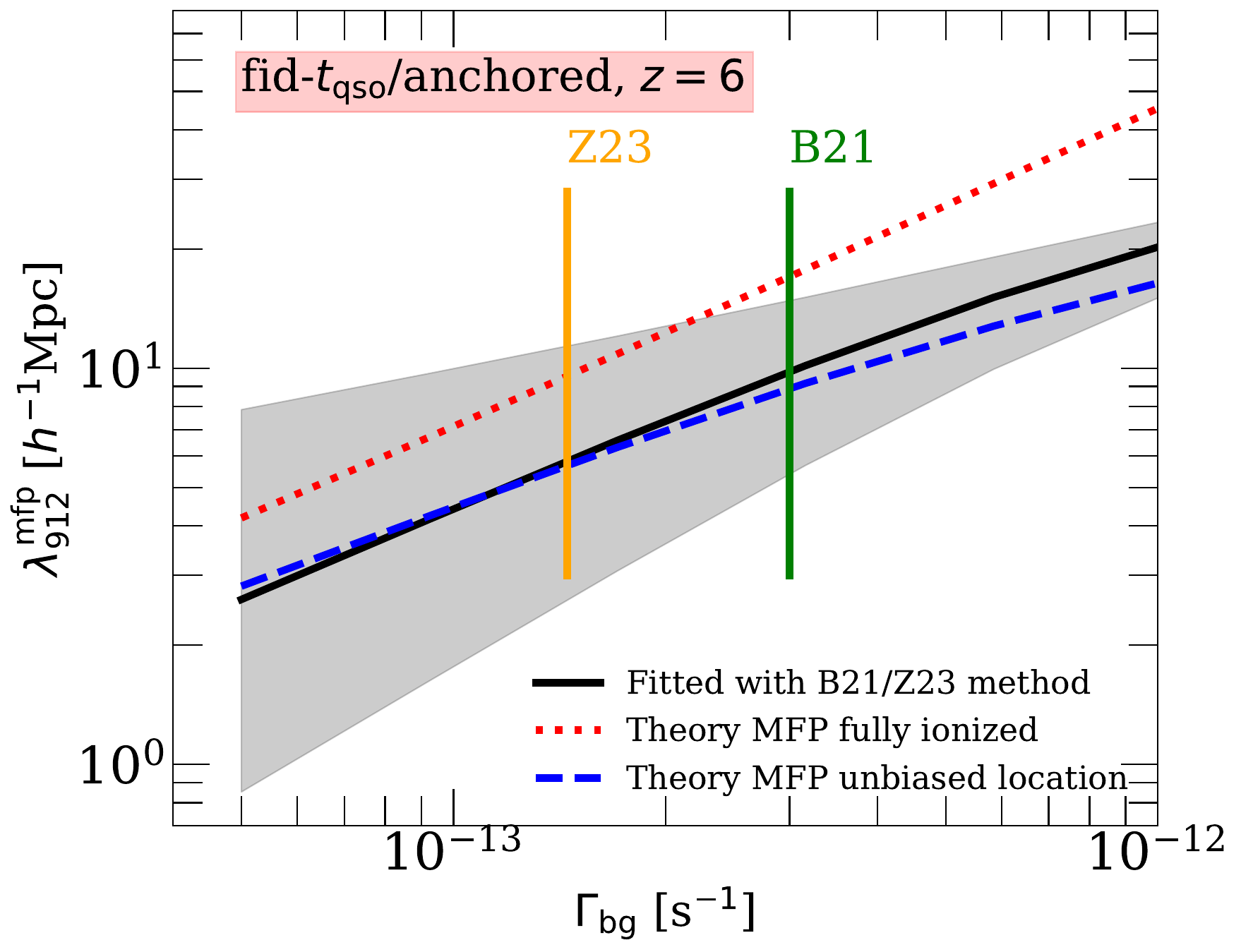}
    \caption{The same quantities as in Fig.~\ref{fig:main_results}, but vs. $\Gamma_{\rm bg}$ for our fiducial model at $z = 6$ (see text).  }
    \label{fig:gammabg}
\end{figure}

\begin{figure}
\centering
    \includegraphics[scale=0.3]{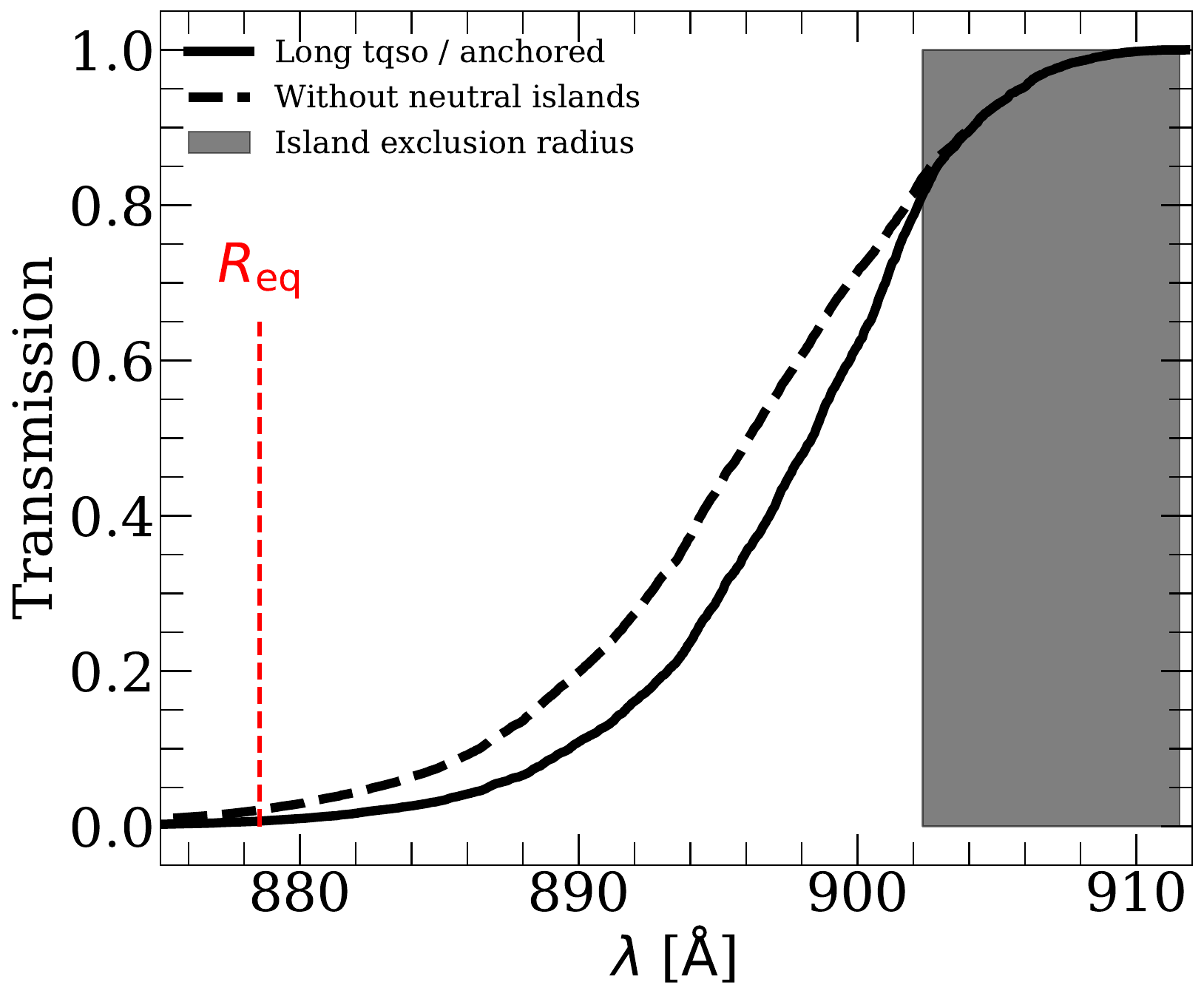}
    \caption{{The effect of neutral islands on a simulated composite QSO spectrum. The solid black curve shows the composite from our long-$t_{\rm qso}$/anchored model, while the dashed-black curve corresponds to a variation of this model without neutral islands. The gray shading corresponds to the exclusion radius from the QSO within which neutral islands cannot reside (see main text), and the vertical dashed-red line shows $R_{\rm eq}$. The composites are nearly identical inside of the exclusion radius. But even in this model, which exhibits the smallest effects from neutral islands among our models, the islands have a significant effect on the shape of the composite.   }}
    \label{fig:islands_effect}
\end{figure}

\begin{figure}
    \centering
    \includegraphics[scale=0.17]{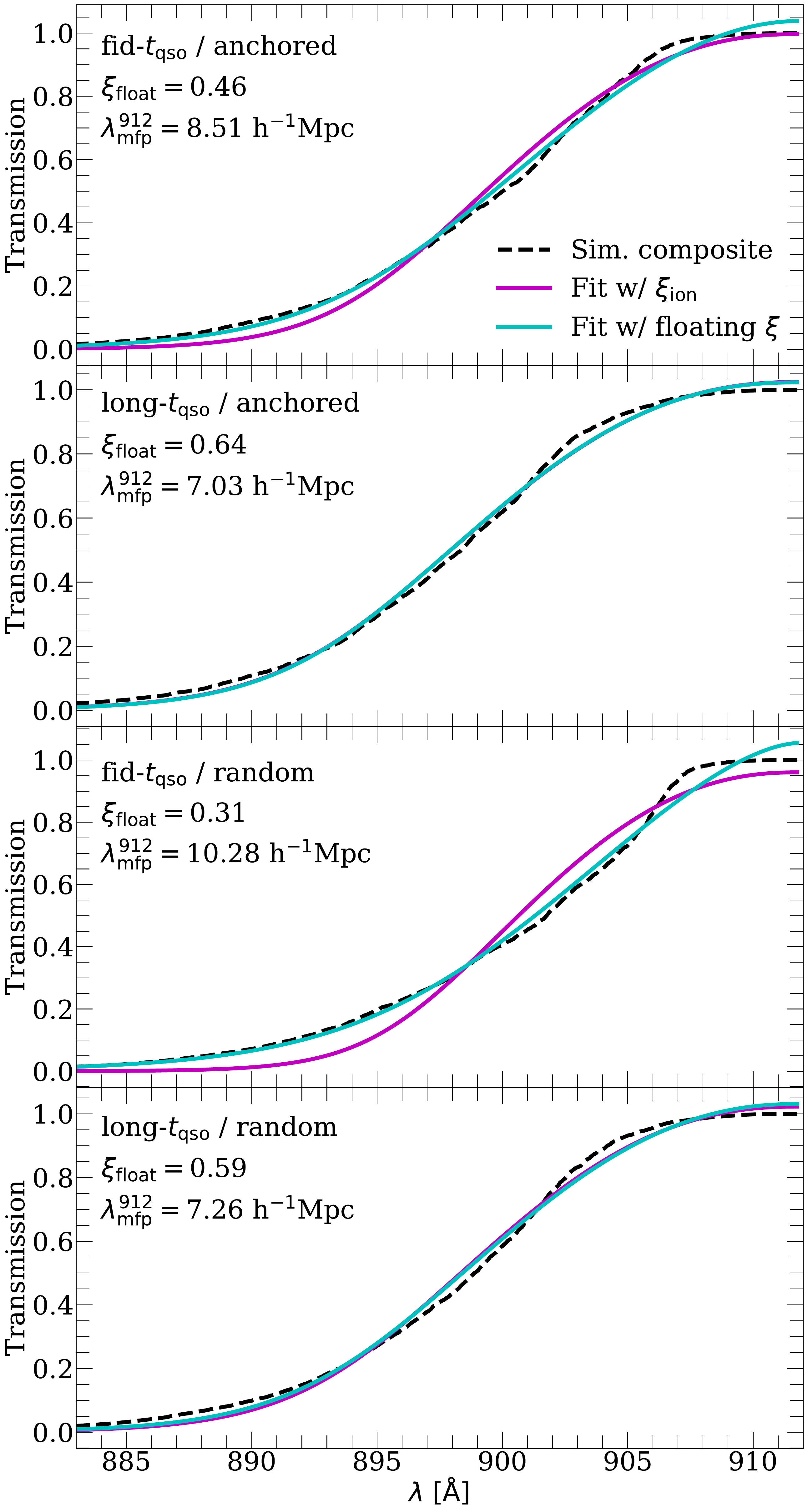}
    \caption{{ Fits to the composite of $1000$ LyC spectra from our simulations (black dashed curves) for each model at $z=6$.  In each panel, the magenta-solid curves show the fits using fixed $\xi_{\rm ion} = 0.63$.  The cyan-solid curves show the fits obtained by letting $\xi$ float in the fitting procedure.  We report these fitted $\xi$ values ($\xi_{\rm float}$) and the corresponding fitted MFP values in each panel. These can be compared to the cases with fixed $\xi_{\rm ion} = 0.63$ provided in the middle column of Table \ref{tab:LyN_effect}, and to the theoretical unbiased-location MFP of $\lambda_{912}^{\rm unbiased} = 5.69~h^{-1}$Mpc.   The fits to some composites are substantially better when $\xi$ is a free parameter, but as $\xi_{\rm float}$ moves lower to accommodate the composite, the inferred MFPs is biased higher with respect to the background MFPs.}}
    \label{fig:lyc_fits}
\end{figure}

Figure~\ref{fig:gammabg} shows results from the fid-$t_{\rm qso}$/anchored model at fixed $z = 6$ with $\Gamma_{\rm bg}$ varying from $5 \times 10^{-14}$ to $2 \times 10^{-12}$ s$^{-1}$.\footnote{Note that for each value of $\Gamma_{\rm bg}$, we recalibrate $\xi_{\rm ion}$ using the procedure described in \S\ref{sec:methods}, to ensure that the B21/Z23 procedure recovers the correct theory MFP when applied to a fully ionized IGM.}  The values adopted in B21 and Z23 are marked by the green and orange vertical lines, respectively.  The agreement between the fit and the unbiased-sightlines MFP holds across a wide range of $\Gamma_{\rm bg}$.  The agreement is particularly good (within $10\%$) at $\Gamma_{\rm bg} < 5 \times 10^{-13}$, which is most realistic for the IGM at $z = 6$. {In any case, inspection of Figs. \ref{fig:main_results} and \ref{fig:gammabg} shows that the fully ionized MFP and the unbiased-location MFP both fall within the fitted values over the range of $\xi \in [0.33,1]$ adopted in B21/Z23 for $\Gamma_{\rm bg} < 2\times10^{-13}$ s$^{-1}$. (And certainly within the range allowed by just the statistical uncertainties in the actual measurements of B21 and Z23, which we do not account for here.) These results lead us to conclude that the B21/Z23 method likely estimates reasonably well the background MFP even in the presence of neutral islands. } 

{So why does the B21/Z23 procedure come so close to the theory MFP in a random patch of the IGM?  Fig. \ref{fig:islands_effect} provides some insight into this question.  It compares simulated composite spectra from our long-$t_{\rm qso}$/anchored model (solid-black) and a variation of this model without any neutral islands (dashed-black). The gray shading corresponds to the exclusion radius from the QSO within which neutral islands cannot reside. We estimate this radius by calculating the distance an $M_{\rm UV}= -27$ QSO I-front could travel in $30$ Myr (the mean of the assumed $t_{\rm qso}$ distribution) through a fully neutral IGM. Note that the two simulated composites are nearly identical in the parts of the spectra corresponding to the exclusion radius. But outside of this radius the neutral islands in the long-$t_{\rm qso}$/anchored model have a significant effect, which is notable because this model by construction exhibits the smallest effects from neutral islands among all of our models. }

{These results suggest that the sensitivity of the B21/Z23 method to the background MFP stems largely from the truncation of the QSO proximity zone by the background opacity, which includes contributions from both the ionized IGM and neutral islands. One consequence of this is that the inferred MFP inevitably has some dependence on the QSO lifetimes, as can be seen in Fig. \ref{fig:main_results}. In fact, in the limit of arbitrarily large $t_{\rm qso}$, the QSO I-fronts will eat away all neutral islands out to a distance $r \lesssim R_{\rm eq}$, where $R_{\rm eq} = 76~h^{-1}$Mpc in our models at $z=6$, shown as the dashed-red line in Fig. \ref{fig:islands_effect}. Hence in that limit, we expect the inferred MFP to approach that of the ionized IGM.  Indeed, we have verified this numerically in a test run with $t_{\rm qso} = 200$ Myr. But for the shorter $t_{\rm qso}$ assumed in our main results, the effect is much milder as already noted.      } 

{An important caveat here is that our calculations do not include the effects of large-scale fluctuations in the cosmic ionizing radiation background, nor do our small simulation volumes sample the highly biased density peaks that likely host high-$z$ QSOs.   There are reasons to expect that including these two effects would not change our main conclusions.  First, B21 tested their fitting method on simulations which used an approximation to RT, the attenuation model of \citet{Davies2016}.  They generated mock composites by tracing sight lines from massive halos in a cosmological simulation with $L = 200~h^{-1}$Mpc. These mock-ups included the local density and $\Gamma$ enhancements from large-scale structure around the host halos -- to the extent that these can be captured in such a box -- as well as the QSO proximity effect. B21 found that their fitting method was able to return the background MFP of the simulation (outside of the proximity zone). Together with the insights provided above, this suggests that any biases from density and $\Gamma$ enhancements are suppressed by the fact that the method probes mainly the opacity on the outskirts of the proximity zone.  }

{
Secondly, Satyavolu et al. in prep. test specifically the effect of the density enhancement using RT reionization simulations populated with QSOs in post-processing. They find that the local density does in general bias the MFP measurements low, but the bias becomes negligible at $z\gtrsim 5.5$, likely due to the competing effect of larger $\Gamma$ fluctuations toward higher redshift. 
Recently, \citet{2023MNRAS.tmp.3051T} used a halo-based analytic model for the LyC opacity to argue that the B21/Z23 method is actually probing the local attenuation scale around the highly biased QSO, which is significantly shorter than the unbiased one. In this scenario, the rapid evolution in the MFP from $z=5$ to $6$ would owe to the increasing bias of the QSO toward higher $z$, and not to an effect of reionization's tail end. The model helpfully provides a way to assess the significance of the bias without the finite box size effects that limit previous analyses.   However, the implications of their results for the B21/Z23 method are difficult to glean because the model does not account for the local $\Gamma$ enhancement from local large-scale structure (which works in the opposite direction), nor the proximity zone.  The latter is especially important because (1) the proximity effect plays a key role in the measurement and (2) as mentioned previously, the edge of the proximity zone may extend beyond the highly biased region close to the QSO.  }

{
Lastly, we return to the question of what happens when $\xi$ is a free parameter. Although the lack of sensitivity in existing data precludes constraining $\xi$, it is worth considering the question for future analyses.   For each panel (model) in Fig.~\ref{fig:lyc_fits}, we plot the simulated composite spectrum at $z=6$ (black-dashed curve) along with two fits using the B21/Z23 method.  The magenta-solid curves show the fits using our fiducial value of $\xi_{\rm ion} = 0.63$, which return the solid-black curves in Fig.~\ref{fig:main_results}.  The solid-cyan curves show the effect of allowing $\xi$ to float in the fitting procedure, such that we obtain a better fit than the fixed-$\xi$ case.  Note that these fitted $\xi$, which we call $\xi_{\rm float}$, tend to be less than $\xi_{\rm ion}$ (with the exception of long-$t_{\rm qso}$/anchored), which is qualitatively consistent with our finding in \S \ref{sec:intuition}.  Moreover, models with a larger effect from islands, e.g. fid-$t_{\rm qso}$ / random, yield a smaller $\xi_{\rm float}$.  }

{
For some cases, the fits with $\xi_{\rm float}$ are obviously better than those with $\xi_{\rm ion}$, although the improvement is not as large as suggested by Fig.~\ref{fig:qso_example}.  This likely owes to the fact that, in the simulations, the QSO itself ionizes some of the islands surrounding it, an effect that is not included in the analytic model. A key result here is that, in cases where $\xi_{\rm float}$ moves significantly lower to accommodate the composite, the inferred MFP is biased high with respect to the unbiased-location MFP of $\lambda_{912}^{\rm unbiased} = 5.69~h^{-1}$Mpc. We take these results as evidence that, as the data improves, and it becomes possible to treat $\xi$ as a free parameter, the method of B21/Z23 may need amending to avoid potential biases in the inferred MFP introduced by neutral islands. Eq. (\ref{eq:kappa_total_r}) provides a possible starting point to that end.    
}

\begin{figure*}
	\includegraphics[width=17.5cm]{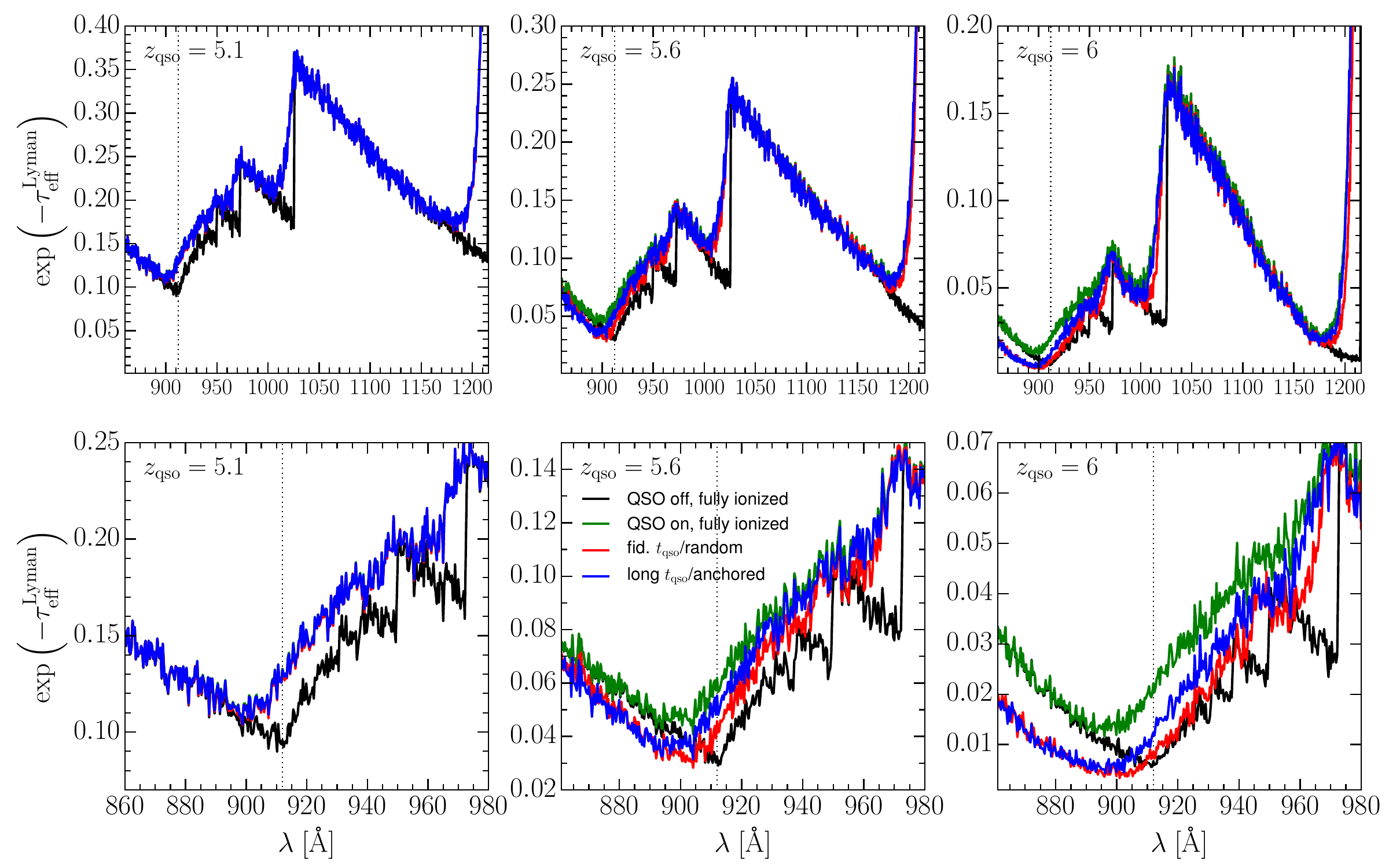}
	\caption{The effect of neutral islands on Lyman-series transmission.  The curves correspond to averaging $1,000$ mock absorption spectra constructed from light cones in our RT simulations. We include Lyman-series lines up to $n=39$. The columns show different $z_{\rm qso}$.  The bottom panels are zoomed-in views of the top panels. Vertical dotted lines show the Lyman limit.  The ionizing background evolution in all models are calibrated to match the observed Ly$\alpha$ forest evolution from \citet{2022MNRAS.514...55B} and \citet{Becker2013}. The black curves assume a fully ionized IGM and neglect the QSO proximity effect, while the green curves include it.  The red and blue curves show two models which include neutral islands from an ongoing reionizaton process between $z=5-6$ (see main text details). Neutral islands suppress transmission through the Lyman-series, introducing substantial differences in the shape of the spectrum. The differences at wavelengths below the Lyman limit, if not modeled adequately, have the potential to affect MFP measurements based on QSO composites.     }
	\label{fig:LyN_transmission}
\end{figure*}

 \section{Lyman series absorption}
\label{sec:LyN}
We  now consider the effect of neutral islands on high-$n$ Lyman series absorption, which manifests in the $\exp\left( -\tau^{\rm Lyman}_{\rm eff} \right)$ term of equation (\ref{eq:total_fobs}).  The models used by B21 and Z23 for this term are obtained from an optically thin hydrodynamics simulation that does not include neutral islands. The question we seek to address here is whether neutral islands could change the shape of $\exp\left( -\tau^{\rm Lyman}_{\rm eff} \right)$, blue-ward of the Lyman limit, enough to bias the inferred MFP in the analyses of B21/Z23.

We model Lyman-series absorption using light cones extracted from a hydrodynamics simulation run with a modified version of the \citet{Trac2004} Eulerian code.  The simulation used $N_{\rm gas}=N_{\rm dm} = 1024^3$ gas cells and dark matter particles in a periodic box with side $L = 20~h^{-1}$Mpc and was run to $z=3$. Skewers are traced through the volume from random positions and random locations. We construct light cones down to $z=3$ by concatenating skewers from neighboring redshift snaphots with spacing $\Delta z = 0.1$ and utilizing periodic boundary conditions.  In what follows, we consider the same models for QSO lifetime and environment as in the last section.  We graft neutral islands to the light cones using the method described in \S \ref{sec:methods}, but this time we extend the neutral island light cone through the completion of reionization. 
 
 To calibrate the models, the ionized neutral hydrogen densities along the lightcones are re-scaled (assuming photoionization equilibrium) to match the observed Ly$\alpha$ forest mean flux evolution using a procedure similar to that in B21/Z23. The observed mean fluxes at $z\geq 5.2$ and $z \leq 4$ are taken from \citet{2022MNRAS.514...55B} and \citet{Becker2013}, respectively, and a power law evolution of $\tau_{\rm eff} = 1.56 [(1 + z )/5.75]^{4}$ is assumed in-between.  The re-scaling is done separately for scenarios with a fully ionized IGM, and for our model with neutral islands.  Finally, we perform 1D RT on the light cones and the output is used to construct mock absorption spectra. We include the Lyman series up to $n=39$ using line data from the National Institute of Standards and Technology. 

Figure \ref{fig:LyN_transmission} shows Lyman-series transmission curves obtained by averaging $1,000$ mock spectra.\footnote{Our curves exhibit more fluctuation than in Fig. 5 of B21 because we take a different approach to calculating the mean transmission, amenable to incoporating RT on our light cones. We compute the transmission for each light cone and then average at the end, whereas B21 computed $\tau^j_{\rm eff}(z_j)$, for transition $j$, from skewers traced through each simulation snapshot volume, and took the observed transmission as $\exp(- \tau^{\rm Lyman}_{\rm eff})(\lambda_{\rm obs}) = \exp[-\sum_j \tau^j_{\rm eff}(z_j)]$, with $(1+z_j)\lambda_j = \lambda_{\rm obs}$, where $\lambda_j$ is the transition wavelength.} (Note that these curves do not include LyC absorption, which would alter their shapes at $\lambda < 912$ \AA.)  The columns correspond to different QSO redshifts while the bottom row shows a zoomed-in view of the top row, covering wavelengths of particular interest for this discussion.  The vertical dotted lines mark the Lyman limit.

The black curves show the transmission in a fully ionized IGM and with no QSO proximity effect, while the green curves show the case when the QSO is on.  Note that, for the $t_{\rm qso}\geq 1$ Myr assumed here, the green curves are insensitive to the particular choice of $t_{\rm qso}$ owing to the line-of-sight effect described in e.g. \citet{White2003}.  The red and blue curves correspond to the fid-$t_{\rm qso}$/random and long-$t_{\rm qso}$/anchored models from the last section.  For clarity, we show only these two because they have the most contrasting assumptions about the presence of neutral islands near the QSOs. In the former, $t_{\rm qso}$ are relatively short and the QSO may turn on in a full neutral region; in the latter, the $t_{\rm qso}$ are longer and the QSO turns on in an already large ionized bubble during reionization.  Note that all models agree in the Ly$\alpha$ forest portions of the spectra by construction from the calibration described above.  

The key takeaway from Figure \ref{fig:LyN_transmission} is that there are significant changes in the Lyman-series transmission, over the wavelength range shown in the bottom panels, owing to the presence of neutral islands in the $z>5$ IGM.  This is most apparent in the $z_{\rm qso}=6$ panels, where the two models with neutral islands are suppressed by $60-100$\% relative to the fully ionized scenario.  The prominent suppression blue-ward of the Lyman limit comes from the ``raking" of the spectrum by a series of closely spaced Lyman-series lines when the redshifted light encounters a neutral island (which have typical sizes of $1-10$ cMpc).  The similarity of the blue and red curves shows that the suppression is insensitive to assumptions about the QSO lifetime and environment at $\lambda < 900$ \AA.  A photon emitted by a $z_{\rm qso} = 6$ QSO with wavelength $\lambda = 900(880)$\AA\ must redshift by $\Delta z = 0.09(0.24)$ before entering the high-$n$ Lyman-series resonances, which corresponds to a travel distance of 27(73) $h^{-1}$cMpc. At these distances and greater the IGM properties are less affected by the QSO radiation.  Note also that the degree of suppression relative to the fully ionized model evolves during the tail of reionization and is absent by $z_{\rm qso}=5.1$. (All of the curves overlap in this case.) It seems likely that such behavior in $\exp\left( -\tau^{\rm Lyman}_{\rm eff} \right)$ mainly probes the neutral fraction along the light cone, so its evolution and shape may turn out to provide a useful empirical constraint on reionization.  In a forthcoming paper we will discuss this possibility in detail (D'Aloisio et al. in prep.).   

The chief question here is whether the shape differences introduced by neutral islands, if not modeled properly, could spoil the measured LyC mean free path in the method of B21/Z23.  To address this, we amend the procedure of the last section to include Lyman-series absorption. 
 From $1,000$ light cones we construct mock composite spectra including LyC and Lyman-series absorption, $\langle \exp(-\tau_{\rm Lyman}) \exp(-\tau_{\rm LyC}) \rangle$.  One composite is created for each of our four models, which span a range of assumptions about $t_{\rm qso}$ and the geometry of neutral islands around the QSOs (as described in the last section). We then fit the composites with equation (\ref{eq:total_fobs}) using the model of B21/Z23 for $\exp(-\tau^{\rm LyC}_{\rm eff})$ and our own version of $\exp(-\tau^{\rm Lyman}_{\rm eff})$ constructed from optically thin RT simulations, i.e. not including neutral islands.  For the latter we use the ``QSO on, fully ionized" models shown as the green curves in Figure \ref{fig:LyN_transmission}, but we increase the number of light cones in the composites to $4,000$ to reduce the noise from cosmological fluctuations in the fitting function.  For brevity we consider only $z_{\rm qso}=6$ in this section. Following the procedure in \S \ref{sec:methods}, we re-calibrated $\xi_{\rm ion}$ but, as expected, the value did not change significantly.  Again, the free parameters of the fits are the normalization, $f_{\rm 912}$, and the MFP.     

\begin{figure}
	\includegraphics[width=8cm]{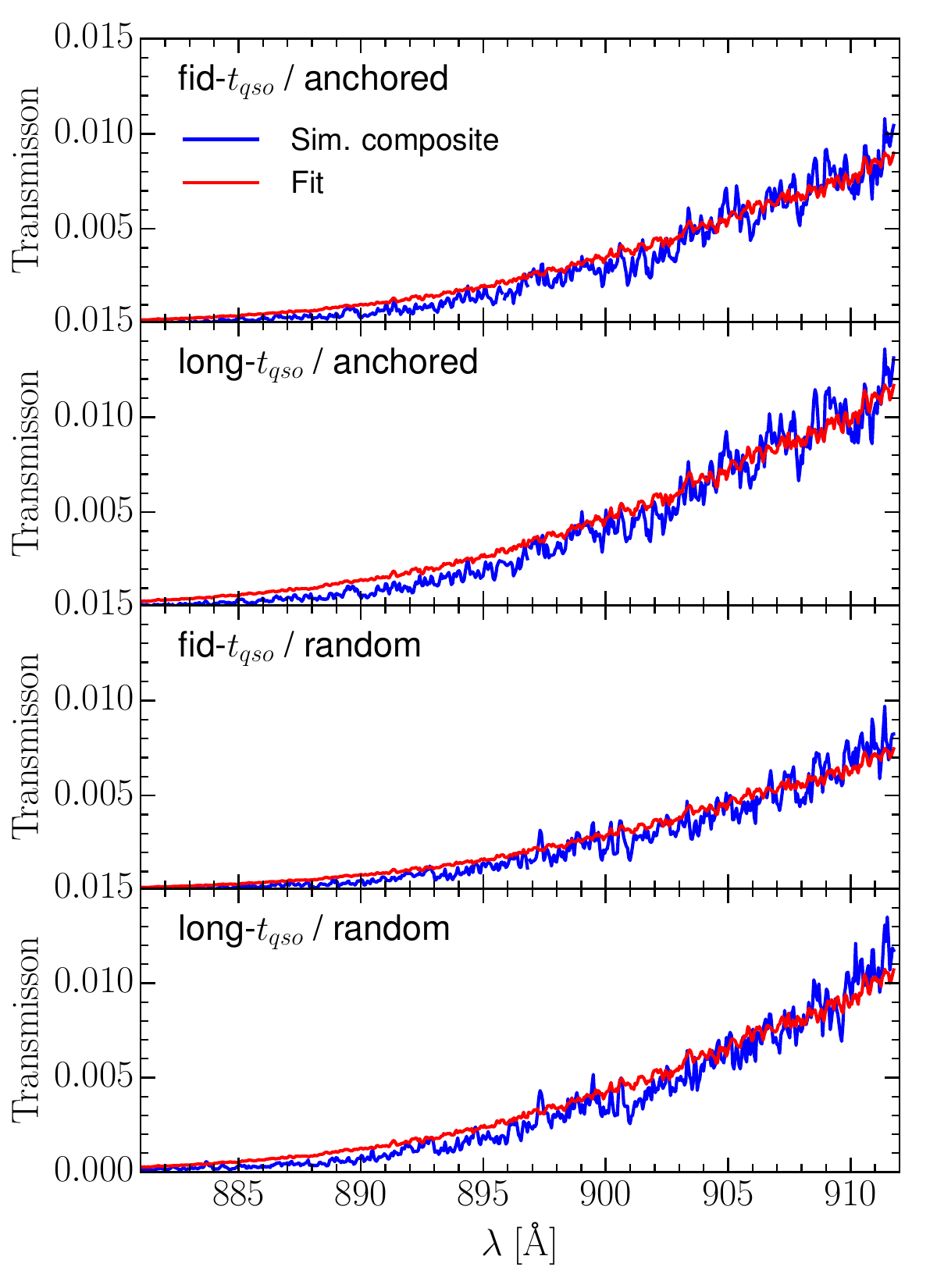}
	\caption{Testing the effects of neutral islands on fits to mock absorption spectra composites with $z_{\rm qso} = 6$. The blue curves show the composites, which are obtained by averaging $1,000$ simulated absorption spectra.  The mocks include LyC and Lyman-series absorption, neutral islands, and the QSO proximity effect (see main text for a description of the four models).  The red curves show fits obtained with a method similar to that of B21/Z23.  The fitting function, $f_{\rm 912} \exp(-\tau^{\rm Lyman}_{\rm eff}) \exp(-\tau^{\rm LyC}_{\rm eff})$, applies Eqs. (\ref{eq:taueff_LyC}) and (\ref{eq:kappa_scaling2}) for the LyC factor, and a model for the Lyman-series term that does not include neutral islands. While some shape differences between the fits and mocks are evident, we find that most of the neutral island effect on $\exp(-\tau^{\rm Lyman}_{\rm eff})$ is absorbed into the normalization, $f_{\rm 912}$.  }
	\label{fig:Fits}
\end{figure}

 The results of the fits at $z_{\rm qso}=6$ are shown in Figure \ref{fig:Fits} for all four models. The blue curves show the mock composite spectra while the red curves show the best fits. There are subtle differences in the shapes of the fitting functions compared to the mock composites, but they would certainly be undetectable in the current data.  Figure \ref{fig:Fits} suggests that the effect of neutral islands on the Lyman-series transmission can be almost entirely absorbed into the normalization, which is in practice embedded in B21's $f^{\rm SED}_\lambda$.  
 
 The best-fit values of the $z=6$ MFP are shown in the right-most column of Table \ref{tab:LyN_effect}.  For reference, the middle column provides the best-fit values from the analysis in the last section, which neglects the Lyman-series absorption in both the mock data and in the fits.  First, note that there were only modest changes in the fitted values of $\lesssim 10$\%, with the exception of the fid-$t_{\rm qso}$ / random model, which changed by $30$\%.  The values should all be compared to the unbiased-location MFP of $5.69~h^{-1}$Mpc from Fig. \ref{fig:main_results}. Our conclusion from the last section still holds: the B21/Z23 procedure yields MFP values close to what we have called the ``unbiased" MFP in an average patch of the IGM far from the QSO. In fact, all of the newly fitted values agree at the $\lesssim 20\%$ level with the unbiased MFP.  
 
 Based on these results we conclude that neutral islands are unlikely to affect the Lyman-series absorption in a way that creates a serious issue for the B21/Z23 MFP measurement method. It appears that the effects at rest-frame $\lambda < 912$ \AA, if present, can be absorbed into the normalization of the spectrum, which B21/Z23 treat as a free parameter, and which does not affect significantly the inferred MFP. 

\begin{table}
\begin{center}
\begin{tabular}{c | c | c }
 Model &  $\lambda^{\rm mfp}_{912}$  (LyC only) & $\lambda^{\rm mfp}_{912}$ (LyC $+$ Ly-series)  \\ [0.5ex]
 \hline
fid-$t_{\rm qso}$~/~anchored  & 5.85 & 6.35 \\ 
 \hline
 long-$t_{\rm qso}$~/~anchored  & 7.11  & 6.66 \\
 \hline
  fid-$t_{\rm qso}$~/~random  & 4.79  & 6.30 \\
 \hline
  long-$t_{\rm qso}$~/~random  & 6.68  & 6.46 \\
  
 \hline
\end{tabular}
\end{center}
\caption{Best fit values for the $z=6$ mean free path in comoving $h^{-1}$Mpc. The last column shows results from our tests including both LyC and Lyman-series absorption, with corresponding fits shown in Fig. \ref{fig:Fits}. For reference, the third column gives the MFP values from our LyC-only analysis in \S \ref{sec:results}. Also for reference, the theoretical unbiased-location MFP at $z=6$ is 5.69$~h^{-1}$Mpc.}
\label{tab:LyN_effect}
\end{table}

\section{Conclusion}
\label{sec:conc} 

{We have studied the effect of neutral islands on the method of B21/Z23 for measuring the $z > 5$ MFP.  We employed a two-pronged approach, using a simple analytic model as well as detailed RT simulations. Testing the B21/Z23 method on mock absorption spectra constructed from our RT simulations, we found that the inferred MFP depends modestly ($< 50 \%$ level) on the QSO environments and lifetimes. Our main finding is that the MFP inferred by the B21/Z23 method would likely estimate reasonably well the background MFP even if neutral islands were present in the composite QSO spectra. (Here ``background MFP" refers to the attenuation scale of the spatially averaged IGM, which includes opacity from both the ionized gas and neutral islands.) Although B21/Z23 held $\xi$ fixed, future data may allow simultaneously constraining the MFP and $\xi$. When $\xi$ is a free parameter, our results suggest that neutral islands will tend to bias the inferred MFP higher than the background value. In this case, the B21/Z23 model may need amending to avoid the potential bias. Eq. (\ref{eq:kappa_total_r}) provides a starting point to that end.}  

We also explored the effects of neutral islands on Lyman-series transmission including absorption lines up to $n=39$. Such effects are relevant to the MFP measurements because they can alter the shapes of the composite spectra blue-ward of rest-frame 912 \AA, relative to models constructed from optically thin simulations.   We found that neutral islands introduce substantial changes to the overall shape of the Lyman-series transmission. At rest-frame $\lambda < 912$ \AA, there can be significant suppression in the transmission (by up to a factor of 2 for $z_{\rm qso} = 6$ in a plausbile reionization scenario) owing to absorption by many closely spaced absorption lines as the light redshifts into resonance while passing through neutral islands.  However, as long as fits are confined to $\lambda < 912$ \AA, we found that these effects are almost entirely degenerate with the fit normalization, which does not affect the inferred MFP value.  We conclude that the B21/Z23 measurements are likely robust to neutral island effects on the Lyman-series transmission. 

An important caveat is that the small simulation volumes used in this work almost certainly do not capture the density structure of the IGM surrounding QSOs, which may reside in extremely rare density peaks.  We also do not account for the possible large-scale enhancement of the ionizing background from the clustering of galaxies near the QSO hosts \citep[e.g.][]{2020MNRAS.494.2937D}.  Future studies should explore whether these omissions modify the conclusions drawn here.

  Our work motivates efforts to constrain empirically the influence of neutral islands on the MFP measurements.   For example, by analyzing the Ly$\alpha$ and Ly$\beta$ forests, it may be possible to place conservative limits on the minimum distance at which a neutral island could exist from the QSOs in the analysis.  Given such constraints, a model for the ionized bubble size distribution around QSOs could be employed with Eq. (\ref{eq:kappa_total_r}) to incorporate neutral islands into the model of B21/Z23. As a summary statistic, the MFP is powerful because it affords intuition about the physical scales at play at the epoch of measurement. {But given the untapped information in the spectra pointed out here, the pursuit of alternative ways of using the data to test reionization models might prove worthwhile.}  

\section*{Acknowledgments}
We thank Matt McQuinn for providing the cosmological skewers used in \S \ref{sec:results}, and Hy Trac for providing the hydrodynamics code that was used to generate the skewers in \S \ref{sec:LyN}. We thank Sindhu Satyavolu, Girish Kulkarni, and Martin Haehnelt for useful discussions, and J. Xavier Prochaska for helpful comments on this manuscript. A.D. and C.C. were supported by NASA 19-ATP19-0191, NSF AST-2045600, and JWST-AR-02608.001-A. All computations were performed with NSF XSEDE allocation TG-PHY210041 and the NASA HEC Program through the NAS Division at Ames Research Center. 

%\vspace{-0.5cm}
%%%%%%%%%%%%%%%%%%%%%%%%%%%%%%%%%%%%%%%%%%%%%%%%%%
\section*{Data Availability}

The data discussed in this article will be shared upon reasonable request to the corresponding author.  

%\vspace{-0.5cm}



\bibliographystyle{aasjournal}
\bibliography{references} % if your bibtex file is called example.bib

%\begin{thebibliography}{99}
%\end{thebibliography}

\end{document}